\documentclass{PoS}

\usepackage{graphicx}
\usepackage{setspace}
\usepackage{amsmath}
\usepackage{color}

\def\be{\begin{equation}}
\def\ee{\end{equation}}
\def\bea{\begin{eqnarray}}
\def\eea{\end{eqnarray}}

\newcommand{\calo}{{\cal O}}

\newcommand{\Tr}{{\rm Tr}\,}

\newcommand{\ax}{a_x}
\newcommand{\axd}{a_x^\dagger}
\newcommand{\bx}{b_x}
\newcommand{\bxd}{b_x^\dagger}

\title{Hybrid Monte-Carlo simulation of interacting 
tight-binding model of graphene}

\ShortTitle{HMC simulation of interacting tight-binding model of graphene}

\author{\speaker{Dominik Smith}%
	\\
        Theoriezentrum, Institut f\"ur Kernphysik, TU Darmstadt,
        64289 Darmstadt, Germany\\
        E-mail: \email{smith@theorie.ikp.physik.tu-darmstadt.de}}

\author{Lorenz von Smekal\\
        Theoriezentrum, Institut f\"ur Kernphysik, TU Darmstadt,
        64289 Darmstadt, Germany\\
        Institut f\"ur Theoretische Physik, Justus-Liebig-Universit\"at,
        35392 Giessen, Germany\\
        E-mail: \email{lorenz.smekal@physik.tu-darmstadt.de}}

\abstract{In this work, results are presented of Hybrid-Monte-Carlo
simulations of the tight-binding Hamiltonian of graphene, coupled to an instantaneous
long-range two-body potential which is modeled by a Hubbard-Stratonovich auxiliary field.
We present an investigation of the spontaneous breaking of the sublattice symmetry,
which corresponds to a phase transition from a conducting to an insulating phase and which occurs
when the effective fine-structure constant $\alpha$ of the system crosses above a
certain threshold $\alpha_C$. Qualitative comparisons to earlier works on the subject
(which used larger system sizes and higher statistics) are made and it is 
established that $\alpha_C$ is of a plausible magnitude in our simulations. Also, 
we discuss differences between simulations using compact and non-compact variants
of the Hubbard field and present a quantitative comparison of distinct 
discretization schemes of the Euclidean time-like dimension in the Fermion operator.}

\FullConference{31st International Symposium on Lattice Field Theory - LATTICE 2013\\
		July 29 - August 3, 2013\\
		Mainz, Germany}

\begin{document}
\section{Introduction}
In recent years, much interest has arisen in the properties of graphene, a one atom thick sheet of carbon atoms
arranged on a hexagonal "honeycomb" lattice. It has become clear that such a simple
structure generates a magnitude of unusual quantum effects, 
which not only make graphene a promising candidate for a wide range of technological applications but also a great
model system to study processes commonly associated with high-energy
physics.\footnote{For an extensive but
far from exhaustive review of graphene phenomenology see
Refs. \cite{CastroNeto:2009zzKotov}.}
The later stems from the fact that the tight-binding Hamiltonian which describes electrons 
in the $\pi$-orbitals of the carbon atoms (with additional terms
describing electromagnetic two-body interactions) is well approximated by a variant of 
Quantum Electrodynamics in $2+1$ dimensions in the limit of low energies (see e.g. Ref. \cite{Gusynin:2007ix}).
In this limit, electronic quasi-particles behave
as massless Dirac particles, with a relativistic dispersion relation. In contrast to QED however
the interacting theory is strongly coupled, since the small Fermi velocity of $v_F\approx c/300$ of the electrons
generates an effective fine structure constant which is $\alpha \approx 300/137 \approx 2.2$. 

It is this surprising connection to relativistic field-theory which drove the high-energy physics community
to take interest in graphene. The strongly coupled nature of the interacting system motivated the 
utilization of established non-perturbative methods such as the simulation in
discretized spacetimes. One of the main questions in such studies, 
which is of direct consequence to technological applications, is whether or not
there exists a band-gap in the interacting system which is absent in a pure tight-binding model. This corresponds
to a spontaneous breaking of the symmetry under exchange of the two triangular sublattices of the graphene sheet and is mapped onto chiral-symmetry breaking in the low-energy effective theory. 
Moreover, graphene allows for a tuning of the effective coupling constant, by affixing
the sheet to a substrate which generates dielectric screening. 
It is thus of interest whether a phase-transition
from a conducting to an insulating phase takes place for some value of $\alpha$, and whether the
critical value lies in a range which can be experimentally realized (suspended graphene giving an upper bound).
Since recent experiments provide evidence that graphene in vacuum is in fact a conductor 
\cite{Elias:2011xvMayorov},
it is important to establish model calculations which match the observation. 

Early attempts at simulating graphene on the lattice studied the low-energy limit only, since the
application of QCD methods is most direct here. Most prominently, Refs. \cite{Drut:2008rg}
simulated the low-energy theory using staggered Fermions, whereas Ref. \cite{Armour:2009vj} investigated
a variant of the \emph{Thirring} model which has many similarities with $\textrm{QED}_{2+1}$. Both find
a phase transition to a gapped phase for $\alpha > \alpha_C \approx 1$, which is well within the physical region. 
More recently, a path integral formulation of the partition function was derived
directly from the tight-binding theory, which preserves the hexagonal structure of graphene 
and employs unphysical discretization in the time dimension only \cite{Brower:2012zd},
with the spatial lattice spacing as a free parameter that can be taken from experiment.
In this formulation, interactions are modeled by a non-local potential
which is generated by a Hubbard-Stratonovich field. 
It is immediately clear that this has many advantages. Not only does it alleviate issues concerning 
parameter matching, it also opens up new opportunities to study physics beyond low energies,
such as the effect of interactions on the neck-disrupting Lifshitz transition, which
occurs in the tight-binding theory at the Van Hove singularity at finite density \cite{Dietz:2013sga}.
Recently, two distinct simulations of the hexagonal lattice have been conducted: One where an unscreened
Coulomb potential is modeled by gauge links \cite{Buividovich:2012nx} and another where the instantaneous
two-body potential was used \cite{Ulybyshev:2013swa}. The later work chose a form of the potential which accounts
for additional screening by electrons in the $\sigma$-orbitals, which was calculated within a constrained
random phase approximation (cRPA) in Ref. \cite{Wehling:2011df}. While simulations with the unscreened potential 
yielded a prediction for $\alpha_C$ consistent with the previous simulations of the low-energy theory, Ref.
\cite{Ulybyshev:2013swa} found evidence that screening is a mechanism which can move $\alpha_C$ to larger values,
possibly outside of the physically accessible region.

In this work, first results of a very recently completed Hybrid-Monte-Carlo code are presented, which simulates the
hexagonal theory with a non-local potential as in Refs. \cite{Brower:2012zd} entirely on GPUs. 
We present an investigation of the conductor-insulator phase transition on a rectangular graphene sheet of $N_x=N_y=6$ 
(counting coordinates in one triangular sub-lattice) and show that,
when properly choosing the exact form of the potential, we find an $\alpha_C$ which is of a plausible magnitude
compared to the previous works. We also conduct an investigation of the discretization errors of the second order
discretization scheme developed in Ref. \cite{Ulybyshev:2013swa} and discuss distinct treatments of the
Hubbard field. Our long-term goal is to obtain a precise prediction
for $\alpha_C$ and to investigate the physics of the Van Hove singularity
for the interacting case.

\section{The path-integral and Hybrid-Monte-Carlo}
 
A detailed derivation of the path integral formulation of the partition function of the interacting tight-binding model
is presented in Refs. \cite{Brower:2012zd}. We only review the crucial steps here. 
The starting point is the Hamiltonian of the model in second quantized form, which is 
\begin{equation}
H = H_{tb}+H_c+H_m = \sum_{\langle x,y \rangle,s}(-\kappa)(a_{x,s}^\dagger a_{y,s}+a_{y,s}^\dagger a_{x,s}) 
+ \sum_{x,y}\, q_x V_{xy} q_y +  \sum_{x}m_S (a_{x,+1}^\dagger a_{x,+1}+a_{x,-1} a_{x,-1}^\dagger) 
\end{equation}
Here $a_{x,s}^\dagger, a_{x,s}$ are Fermionic creation- and annihilation operators which respect the usual 
anti-commutation relations. $s$ denotes the electron spin and takes the values $\pm 1$. $\kappa\approx 2.8 \textrm{eV}$ 
is the hopping parameter. $m_S=\pm m$ is a ``staggered'' mass, which has a different sign on each sub-lattice and which
is added to serve as a seed for sub-lattice symmetry breaking (simulations are extrapolated to $m \to 0$). The first
sum runs over all pairs of nearest neighbor sites, while the second and third sums run over all pairs of sites and
all sites respectively. The interaction matrix $V_{xy}$ need not be further specified at this point, other than that it
be positive definite.
$q_x=a^\dagger_{x,1}a_{x,1}+a^\dagger_{x,-1}a_{x,-1}-1$ is the charge operator where the constant is added to make the
system neutral at half filling. To prevent a type of Fermion sign problem from occurring later,
two transformations are made: To introduce new operators
$b_x^\dagger=a_{x,-1}~,~b_x=a_{x,-1}^\dagger$ and flip the sign of these on one sub-lattice.

The functional integral for $Z=\Tr e^{-\beta H}$ (here $\beta$ is $1/k_B T$) can now be derived using
a \emph{coherent state} formalism.\footnote{The temperature is that of
  the gas of electronic quasi-particles only 
and is not equal to the physical temperature of
the graphene sheet.} 
The essential step is to factor the exponential into $N_t$ terms 
$~e^{-\beta H}=e^{-\delta H}\,e^{-\delta H}\ldots e^{-\delta H},$ ($\delta=\beta/N_t$)
and to insert between them complete sets of Fermionic coherent states
$
\langle\psi_k,\eta_k|=\langle 0| e^{-\sum_x (\ax \psi_{x,k}^{*}+\bx \eta_{x,k}^{*})}~,
|\psi_k,\eta_k\rangle= e^{-\sum_x (\psi_{x,k}\axd + \eta_{x,k}\bxd)}|0\rangle~.
$
This leads to the expression
\be
\Tr e^{-\beta H}=\int \prod\limits_{t=0}^{N_t-1} \left[\prod\limits_x
        d\psi^{*}_{x,t} \, d\psi_{x,t} \, d\eta^{*}_{x,t} \, d\eta_{x,t} \right]
        ~e^{-\sum_x( \psi^{*}_{x,t+1} \psi_{x,t+1}+\eta^{*}_{x,t+1} \eta_{x,t+1} )} 
        \langle \psi_{t+1},\eta_{t+1}|e^{-\delta H}|\psi_{t},\eta_{t}\rangle~. \label{eq:partfunc}
\ee
Here $\psi_{x,t}$ and $\eta_{x,t}$ are Grassman valued
field variables which replace the ladder operators. 
Anti-periodic boundary conditions in the time-direction
are implied $(\psi_{x,N_t}=-\psi_{x,0}~,~\eta_{x,N_t}=-\eta_{x,0})$. We now
use the identity
$
\langle\psi,\eta|F(a^{\dagger}_\lambda,a_\lambda,b^{\dagger}_\lambda,b_\lambda)|\psi',\eta'\rangle = 
F(\psi^{*}_\lambda,\psi'_\lambda,\eta^{*}_\lambda,\eta'_\lambda)\, 
~e^{\sum_\lambda \psi_\lambda^{*} \psi_\lambda'+\eta_\lambda^{*} \eta_\lambda'},~
$
which uses coherent states to map normal ordered functions of ladder operators to functions of Grassman variables
and obtain 
\begin{align}
\Tr e^{-\beta H} &= \int \prod_{t=0}^{N_t-1} \left[ \prod_x d\psi^{*}_{x,t} \, d\psi_{x,t} \, d\eta^{*}_{x,t} \, d\eta_{x,t} \right]
\exp \Big\{ -\delta  \Big[\sum_{x,y} Q_{x,t+1,t}V_{xy}Q_{y,t+1,t}\notag\\
&-\sum_{\langle x,y \rangle}\kappa
(\psi_{x,t+1}^{*} \psi_{y,t}+\psi_{y,t+1}^{*} \psi_{x,t}+\eta_{y,t+1}^{*} \eta_{x,t}+\eta_{x,t+1}^{*} \eta_{y,t}  )+\sum_x m_S(\psi_{x,t+1}^{*}\psi_{x,t}+\eta_{x,t+1}^{*}\eta_{x,t})\notag\\
&+\sum_x V_{xx}(\psi_{x,t+1}^{*}\psi_{x,t}+\eta_{x,t+1}^{*}\eta_{x,t})\Big] 
-\sum_x\big[ \psi^{*}_{x,t+1} (\psi_{x,t+1}-\psi_{x,t}  )+\eta^{*}_{x,t+1} (\eta_{x,t+1}-\eta_{x,t}) \big] \Big\}~.
\label{eq:partfunc2}
\end{align}
Here we have applied normal ordering to $H$, which leads to the additional term $\sim V_{xx}$ and 
then considered $e^{-\delta H}$ to be normal ordered, which implies
a discretization error $\calo(\beta/N_t)$ (which vanishes for $N_t \to \infty$).
Also, we have defined a charge field $Q_{x,t,t'}=\psi_{x,t}^{*}\psi_{x,t'}-\eta_{x,t}^{*}\eta_{x,t'}$.

Eq. (\ref{eq:partfunc2}) contains fourth powers of $\psi,\eta$. We must get rid of these if we wish to
carry out Gaussian integration and obtain a Fermion determinant, which
is necessary for HMC. 
This is achieved by applying the Hubbard-Stratonovich transformation
\begin{align}
e^{ -\delta \sum_{t=0}^{N_t-1}  \sum_{x,y}Q_{x,t+1,t}V_{xy}Q_{y,t+1,t} }
\sim \int \left[ \prod_{t=0}^{N_t-1} \prod_x \, d\phi_{x,t}\,\right]
e^{-\frac{\delta}{4} \sum_{t=0}^{N_t-1} \sum_{x,y}
\phi_{x,t} V_{xy}^{-1} \phi_{y,t}-i\,\delta
\sum_{t=0}^{N_t-1} \sum_{x} \phi_{x,t} Q_{x,t+1,t}}~, \label{eq:hubbard1}
\end{align}
which eliminates fourth powers, at the expense of introducing an
auxiliary field $\phi_{x,t}$ and 
introducing the inverse of the matrix $V_{xy}$. The constant factor before the integral in Eq. (\ref{eq:hubbard1}) can be dropped. Combining equations (\ref{eq:partfunc2}) and (\ref{eq:hubbard1}) and carrying out Gaussian integration
finally yields
\be
\Tr e^{-\beta H}=\int\left[  \prod_{t=0}^{N_t-1} \prod_x
\, d\phi_{x,t}\right]
e^{-\frac{\delta}{4} \sum_{t=0}^{N_t-1} \sum_{x,y}
\phi_{x,t}V_{xy}^{-1} \phi_{y,t} 
} \left| \det M(\phi) \right|^2~.
\ee
This form is suitable for simulation via HMC. The
determinant can be sampled stochastically by using pseudo-Fermion
sources and thus $\phi$ remains as the only "dynamical" field. 
The matrix $M$ is defined in terms of its components as
\begin{align}
\frac{N_t}{\beta} M_{(x,t)(y,t')} =\frac{N_t}{\beta} \delta_{xy}(\delta_{tt'}-\delta_{t-1,t'})-
\kappa \sum\limits_{\vec{n}} \delta_{y,x+\vec{n}}\delta_{t-1,t'}+ 
m_S \delta_{xy} \delta_{t-1,t'} + 
V_{xx}\delta_{xy}\delta_{t-1,t'}
+i\phi_{x,t}\delta_{xy} \delta_{t-1,t'}~.\label{eq:fermionmatrix}
\end{align}
Here $\vec{n}$ denotes vectors connecting nearest-neighbor sites. Since
$M$ is hermitian, there is no sign problem. 

Simulations based on Eq. (\ref{eq:fermionmatrix}) have a severe problem:
The term $\sim \phi$
makes $\det M(\phi)$ a polynomial of order $\phi^N$ where $N$ is the number
of lattice sites. Thus, rounding errors are 
amplified in an uncontrollable way. As is discussed further in
Section \ref{sec:results}, this indeed make simulations using
(\ref{eq:fermionmatrix}) impossible. The solution was worked out
in Refs. \cite{Brower:2012zd}, where it is shown that one may
replace
\be
\frac{\beta}{N_t}V_{xx}\delta_{xy}\delta_{t-1,t'}+i\frac{\beta}{N_t}\phi_{x,t}\delta_{xy} \delta_{t-1,t'}
\longrightarrow 
e^{ i\frac{\beta}{N_t}\phi_{x,t}} \delta_{xy} \delta_{t-1,t'} ~,
\ee
which transforms $\phi$ into a phase that is additive in $\det M(\phi)$
and thus numerically stable.

It should be noted that that (\ref{eq:fermionmatrix}) is by far not the only possible
form of $M$. As Refs. \cite{Brower:2012zd} discuss,
there is a great deal of freedom in discretizing the time-direction and this can be exploited
to construct improved actions which converge faster to the continuum limit. 
A second order action was presented 
in Ref. \cite{Ulybyshev:2013swa}, which is constructed by factoring $\exp(-\delta H)$
such that the interacting part is split off, and inserting an additional set of
coherent states between the terms:
\begin{align}  
\Tr e^{-\beta H}=&\int \left[\prod_{t=0}^{2N_t-1} \prod_x
d\psi^{*}_{x,t} \, d\psi_{x,t} \, d\eta^{*}_{x,t} \, d\eta_{x,t} \right]
\prod_{t=0}^{N_t-1}~e^{-\sum_x(  \psi^{*}_{x,2t} \psi_{x,2t}+\eta^{*}_{x,2t} \eta_{x,2t}
+\psi^{*}_{x,2t+1} \psi_{x,2t+1}+\eta^{*}_{x,2t+1} \eta_{x,2t+1} )}\notag\\
&\times \langle \psi_{2t},\eta_{2t}|e^{-\delta (H_{tb}+H_m)}|\psi_{2t+1},
\eta_{2t+1}\rangle \langle \psi_{2t+1},\eta_{2t+1}|e^{-\delta H_C}|
\psi_{2t+2},\eta_{2t+2}\rangle~.
\end{align}
Carrying out the calculation in analogy to what was discussed above (using the compact version
of the Hubbard field), one obtains the following Fermion operator:
\be
M_{(x,t)(y,t')}  = \left\{
\begin{array}{lr}
\delta_{xy}(\delta_{tt'}-\delta_{t+1,t'})-\frac{\beta}{N_t} \kappa 
\sum\limits_{\vec{n}} \delta_{y,x+\vec{n}}\delta_{t+1,t'}
+ \frac{\beta}{N_t}m_S \delta_{xy} \delta_{t+1,t'} & : t~\textrm{even}\\
\delta_{xy}\delta_{tt'}- \delta_{xy} \delta_{t+1,t'}
\exp(-i \frac{\beta}{N_t} \phi_{x,(t-1)/2})& : t~\textrm{odd}~ 
\end{array}
\right.\label{eq:fermionmatrix_2nd}
\ee
Note that the Hubbard field now only appears on odd time-slices. It was hypothesized
in Ref. \cite{Ulybyshev:2013swa} that this version of the Fermion operator has 
reduced discretization errors. 

We wish to investigate spontaneous breaking of sub-lattice symmetry. Thus we require
a proper order parameter. The obvious choice is to use the difference of number density
operators on the two sub-lattices (denoted here as $A$ and $B$). 
In the functional integral form, this is expressed as 
\begin{align}
\langle \Delta_N \rangle &=\frac{1}{ZN_t}\int\mathcal{D}\psi\mathcal{D}\psi^*\mathcal{D}\eta\mathcal{D}\eta^*
\Big[ \sum_{X_A,t}\left(\psi^*_{x,t+1}\psi_{x,t}+\eta^*_{x,t+1}\eta_{x,t}\right) - \sum_{X_B,t}\left(\psi^*_{x,t+1}\psi_{x,t}+\eta^*_{x,t+1}\eta_{x,t}\right) \Big]e^{-\beta H} \notag\\
&=\frac{-1}{\beta Z}\int\mathcal{D}\phi
\left[ \frac{\partial}{\partial m} \det\left(M M^\dagger \right)
 \right] e^{-S[\phi]}  
=\frac{-2}{N_t} \sum\limits_{t=0}^{N_t-1}
\langle \sum\limits_{x \in A} M^{-1}_{(x,t+1)(x,t)}-
\sum\limits_{x \in B} M^{-1}_{(x,t+1)(x,t)}\rangle ~.\label{eq:pbp1}
\end{align}
The above is valid for the first order discretization scheme above. For the second order
Fermion operator, one may insert $\widehat{\Delta}_N$ on even time-slices only to
obtain an analogous expression. 

\section{Results}
\label{sec:results}
We have simulated rectangular $N_x=N_y=6$ (base vectors defined in a triangular sub-lattice)
graphene sheets with periodic boundary conditions using both the first and second order
Fermion operators. We chose a potential which is constructed
piecewise: The on-site ($V_{00}$), nearest-neighbor ($V_{01}$), next-nearest neighbor ($V_{02}$)
and crossing term (across the hexagon, $V_{03}$) are the cRPA values
of Ref. \cite{Wehling:2011df}, 
while the long-range part is an unscreened Coulomb potential. 
We account for periodicity by adding one set of eight mirror images,
arranged along the rectangular boundary.
This may be problematic due to the zero mode in the potential. Moreover, we
have chosen $V_{00}=4.15\textrm{eV}$ which is slightly smaller than in Ref. \cite{Wehling:2011df} in a
attempt to match to \cite{Wehling:2011df} after the contribution of
mirror charges is taken into account (probably one should not do this
and use $V_{00}=4.65\textrm{eV}$ directly). 
Also, at this stage, the toroidal structure of the lattice
was not properly taken into account: We took the shortest path that does not cross the boundary
instead of the shortest actual path as the "direct" connection. Addressing these issues is work in
progress.
We chose $\beta=2.0{~\textrm eV^{-1}}$ (as in Ref. \cite{Ulybyshev:2013swa}). 
An effective $\alpha=(300/137)(1/\epsilon)$ is introduced
as a free parameter which implies a rescaling of $V_{xy} \to V_{xy}/\epsilon$.
Our first observation is that the non-compact Hubbard field is indeed problematic. Foremost, we were
not able to observe distinctly non-zero expectation values of the order parameter. We thus use 
the compact version only. The table below shows a comparison of the discretization errors of
the first and second order discretization schemes (labeled "standard" and "improved" in the following).
These were obtained by measuring $\langle \Delta_N \rangle$ on roughly one hundred independent
configurations for each combination of $(\alpha,m,N_t)$ and fitting results obtained for different $N_t\in[12, \ldots, 28]$
to $\langle \Delta_N \rangle= c_1*(1/N_t)+c_2$.
\begin{center}
\vspace{2mm}
\begin{tabular}{|c|c|c|c|c|c|c|}
\hline
$\beta [{\textrm eV^{-1}}]$ & $\alpha$ & $m [{\textrm eV}]$ & $c_{1,\textrm{std}}$ & $c_{1,\textrm{imp}}$ & $c_{2,\textrm{std}}$ 
& $c_{2,\textrm{imp}}$ \\
\hline
$2.0$ & $1.87$ & $0.3$ & $-0.74(8)$ & $-0.81(3)$ & $0.31(1)$ & $0.320(4)$ \\
\hline
$2.0$ & $1.87$ & $0.5$ & $-1.01(7)$ & $-0.98(8)$ & $0.462(8)$ & $0.463(8)$ \\
\hline
$2.0$ & $2.55$ & $0.5$ & $-1.5(1)$ & $-1.3(1)$ & $0.566(3)$ & $0.563(3)$ \\
\hline
$2.0$ & $2.18$ & $0.5$ & $-1.02(5)$ & $-1.4(2)$ & $0.493(5)$ & $0.54(3)$ \\
\hline
\end{tabular}
\vspace{2mm}
\end{center}
These results provide strong evidence that both versions not only converge to the same continuum
limit but are in fact equal for any $N_t$. It thus appears that nothing is to be gained
by using the second order action.

\begin{figure}
\begin{center}
\resizebox{0.49\textwidth}{!}{%
\includegraphics{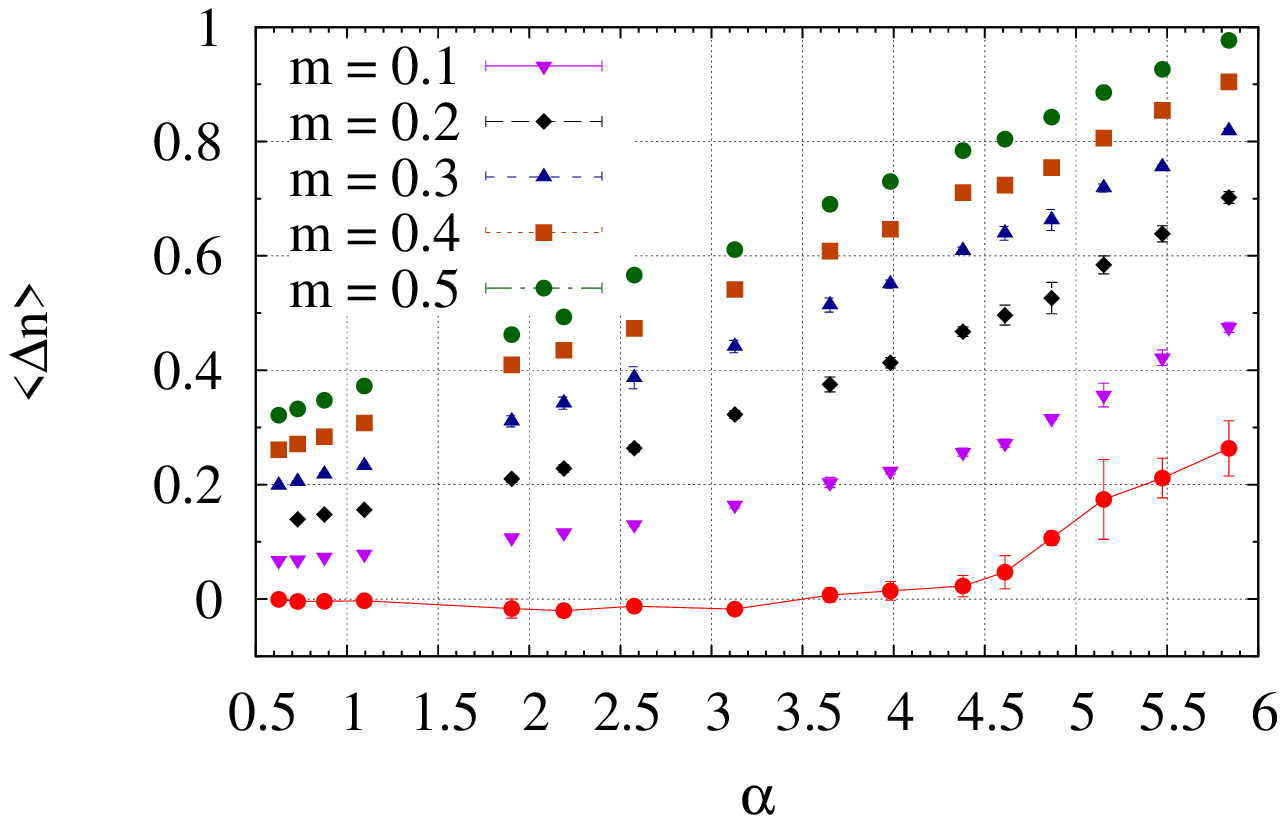}}
\resizebox{0.49\textwidth}{!}{%
\includegraphics{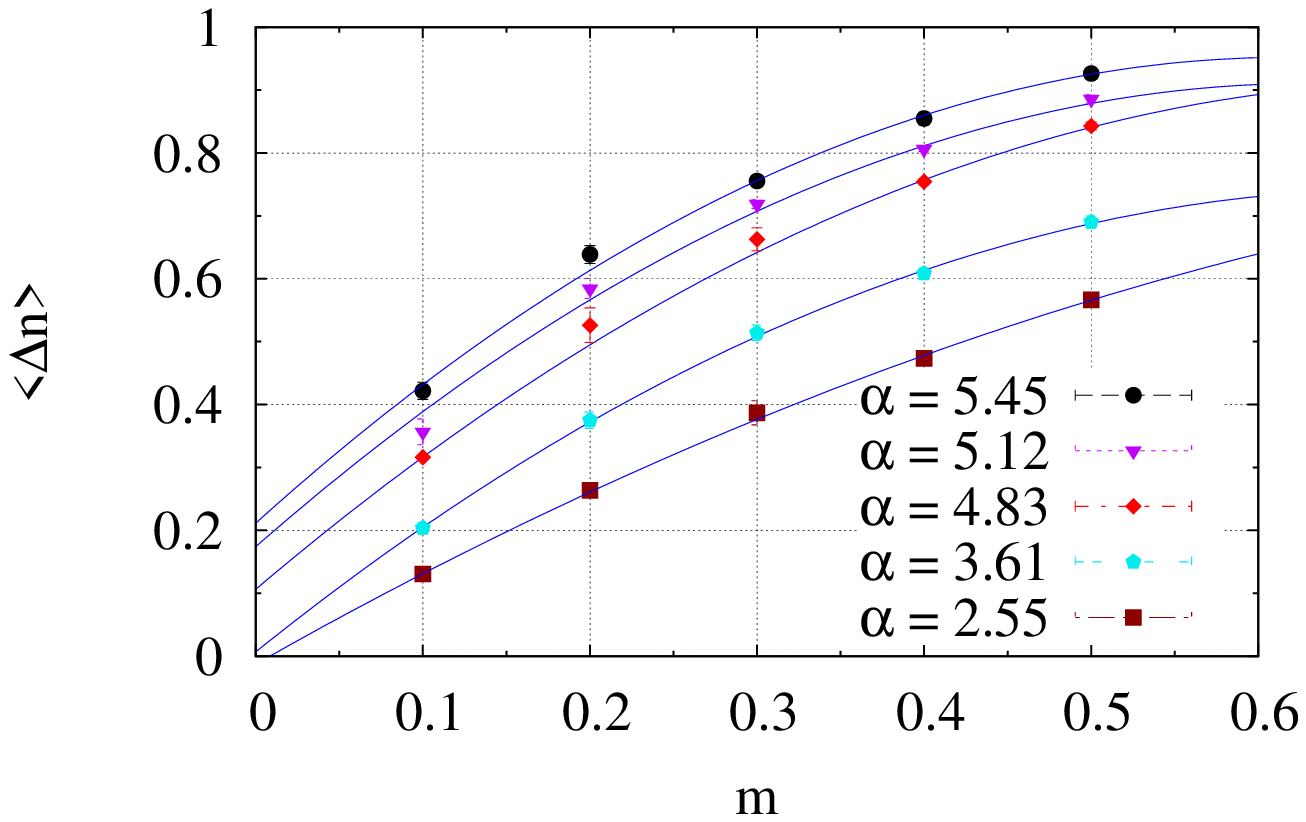}}
\end{center}
\caption{Phase transition of $N_x=N_y=6$ system (left figure). The red line and dots 
represent the $m \to 0$ extrapolated results. The right figure shows the extrapolation for a
few exemplary choices of $\alpha$.  \label{fig:fig2}}
\end{figure}

Subsequently we performed an investigation of the phase transition. We simulated for many
different choices of $\alpha$, where masses were chosen as $m=0.5,0.4,0.3,0.2,0.1\textrm{eV}$ for
each case and simulations were extrapolated to the continuum from $N_t=[12, \ldots, 28]$
for each set of $(\alpha,m)$.
Again roughly one hundred independent measurements were done for each set $(\alpha,m,N_t)$
The continuum results are shown in Fig. \ref{fig:fig2}. We have extrapolated to $m=0$ for each $\alpha$
using $\langle \Delta_N \rangle= c_1 m^2+c_2 m+c_3$. We find that our choice of potential
generates an $\alpha_C \approx 4$.



\section{Conclusions and Outlook}
We have demonstrated here that our code produces qualitatively the expected behavior,
with a phase transition occurring at an $\alpha_C\approx 4$ which is not far from
what was found in Ref. \cite{Ulybyshev:2013swa} ($\alpha_C\approx 3.12$).
We stress that our result represents
a plausibility check, not a prediction. The potential we have chosen differs from
Ref. \cite{Ulybyshev:2013swa} and the system size is much smaller. Moreover, boundary
conditions were not addressed in the same way. Very recently, there has been substantial progress in
this regard: We have implemented a computation of the potential via Fast Fourier Transformation,
which now allows us to simulate at larger volumes. When choosing the same potential as
Ref. \cite{Ulybyshev:2013swa} and properly addressing boundary conditions, it appears we are
now able to reproduce their results also quantitatively. We are thus confident that
our code is now ready for large-scale operation.

\medskip

\leftline{\textbf{Acknowledgements}}

\smallskip
 

We have benefited from discussions with Pavel Buividovich, Maxim
Ulybyshev and the late Mikhail Polikarpov. This work was supported by
the Deutsche Forschungsgemeinschaft within 
SFB 634, by the Helmholtz International Center for FAIR within the
LOEWE initiative of the State of Hesse, and the European Commission,
FP-7-PEOPLE-2009-RG, No. 249203.  All results were obtained using
Nvidia GTX and Tesla graphics cards.

\end{document}